\documentclass[manuscript]{aastex}

\shortauthors{GONG ET AL.} \shorttitle{Precession velocity of
X-sources} \received{2011 May 12} \accepted{}

\begin{document}

\title{The first kinematic determination of million-year precession period of AGNs
}

\author{B.P. Gong\altaffilmark{1},
Y.P. Li\altaffilmark{1},
  H.C. Zhang\altaffilmark{2}
}

\altaffiltext{1}{Department of Physics, Huazhong University of
Science and Technology, Wuhan 430074, China}
\altaffiltext{2}{Department of Physics and Astronomy, The Ohio
University, Athens, OH 45701, USA}


\begin{abstract}
Short precession periods like 164d of SS433  can be well determined
by observations of time scales longer or much longer than the
precession period. However, it doesn't work for sources with
precession periods of millions of years. This paper utilizes the
particular morphologies of X-shaped sources, so that the 3 dimension
kinematics  of lobes can be obtained. Thus, for the first time, the
million-year precession period of X-shaped sources by observer on the Earth
 can be determined elegantly: $(6.1\pm 1.5)$Myr,
$(1.8\pm 0.5)$Myr, and $(3.2\pm 1.2)$Myr for 3C52, 3C223.1 and 4C12.03
 respectively. The result naturally explains the asymmetry
displayed in the morphology of these sources, and the
effect of propagation time on the diversity of morphologies is well
demonstrated.
The precession period may originate from long-term  effects of a binary super-massive black hole system,
 which is a  potential
source of gravitational wave radiation.

\end{abstract}


\section{Introduction}

First discovered in 1974~\citep{HC74} and growing rapidly in recent
years~\citep{Cheung07}, the  peculiar radio morphologies, X-shaped
extragalactic radio sources,  are characterized by two low surface
brightness wings oriented at an angle to the high surface brightness
lobes, giving the total an X-shape. Recently, X-shaped galaxies are being considered as potential transition between
Fanaroff-Riley (FR) type I and II~\citep{Landt10}.

Several formation scenarios have been proposed. One is  the
back-flow of plasma from the active lobes into the
wings~\citep{LW84,cap02,HK11}, with subsequent buoyant expansion. It has
been argued that the expansion of wings is subsonic, and it becomes
untenable for X-shaped sources with wings longer than the active
lobes~\citep{DT02}. The second scenario is the  conical precession
of the jet axis~\citep{parma85,mack94}, which implies a scenario of
ballistic jet motion plus jet precession, predicting
spiral pattern. However, this model requires a specific
accident of the positions at which the source first switched on and its position now. Moreover, it can not explain
the notable asymmetry in 3C223.1~\citep{DT02}.

Other two explanations  have received much attention lately. They
are in agreement that  the wings are the relics left over from a rapid
realignment of a central super-massive black-hole (SMBH) accretion
disk system. The realignment can be a result of  a
relatively recent merger of a super-massive binary black hole
(SMBBH)~\citep{ME02,Bonn} or due to disk-instability~\citep{DT02}.


Nevertheless, in these two scenarios the morphology of X-sources apparently requires a rapid change of jet orientation.
A number of X-sources have companion
galaxies~\citep{DT02}, and the host galaxy of 3C293 shows clearly
 interaction. Moreover, the double-peaked low-ionization
emission lines in the nucleus of a galaxy associated with X-shaped
structure\citep{Zhang07} provide an interesting signature of link
between X-source and SMBBH. Such a link makes X-sources potential
sources of gravitational wave radiation, which leads to a steep
surge of interest in them recently\citep{Komossa03}, although they
have been known for decades.

\begin{figure}[t]
\centering
\epsscale{0.8} \plotone{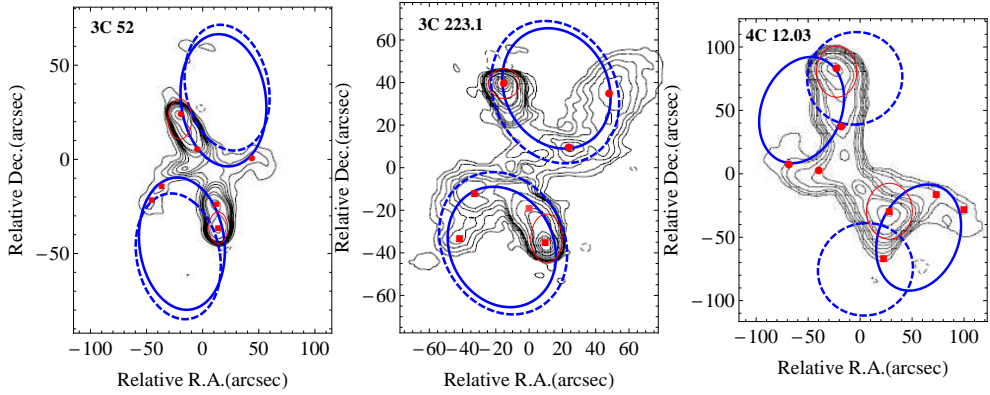}
\caption{\small The observed morphology of three X-shaped sources\citep{Lal07}, and the fitted traces (ellipses) given by the
non-ballistic model.The north is to up, and the east is to left.  The solid ellipses correspond to the best fit parameters of Table~1,
 and the dashed ellipses correspond to 20$\%$ increase in  parameters, $I$, $\lambda$, and $\xi$ (while  holding  others constant) for
 3C52, 3C223.1 and   4C12.03  respectively. The red dots represent components of a morphology which the ellipses try to fit.
The red circles are ``radius" of the lobes through which the error in precession phase is estimated.\label{morphology}}
\end{figure}

In fact, both the jet precession and  reorientation scenarios agree
that the wings are relics of the previous jet, and the lobes are
produced by jet in action. Thus, regardless of the mechanism of
the change of the jet axis,  We establish a coordinate system, in which
the 2-dimensional morphology can be fitted by the simplest geometry.
And together with the constraint imposed  by the simultaneous arrival time of
photons from the south and
north lobes of an X-source, the most collimated  components of
Fig.~$\ref{morphology}$, the 3-dimensional
kinematics of the morphology  can be obtained. Consequently,
the time scale of the formation of the
X-shaped morphology (wing and lobe) can be determined  in a general
manner. Applying to three X-sources of FR-II, 3C52, 3C223.1 and 4C12.03~\citep{ME02,DT02,Lal07} as shown in Fig.~$\ref{morphology}$, the precession periods (by Earth observer) of $(6.1\pm 1.5)$Myr, $(1.8\pm 0.5)$Myr, and $(3.2\pm 1.2)$Myr are obtained  respectively. Throughout this letter we assume a $\Omega_{\Lambda}$=0.73,
$\Omega_m$=0.27 cosmology with $H_0$=71km s$^{-1}$Mpc$^{-1}$. And the red-shift, $z$, of 3C52, 3C223.1 and 4C12.03 are 0.2854, 0.1074 and 0.1570 respectively\citep{ME02,Saripalli09}.

\section{The New Approach}

The non-ballistic model~\citep{Gong08}
has been used to interpret the non-radial jet motions of
AGNs~\citep{Kellermann04,Agudo07,Lister09}, in which a knot can be
produced by a continuous jet interacting with ambient matter in
different directions during the precession of the jet axis. Approximately
equal knot-core separation is expected  when the power of the jet
and matter density of the surrounding medium are unchanged in different
directions.
Such a constant core-knot separation avoids
the specific accident of the positions required by conical precession model.
And  due to the X-shaped morphologies display similar non-radial characteristic
as other AGN sources,   it is conceivable to apply the non-ballistic model to the X-sources.

\begin{figure}[t]
\epsscale{0.8} \plotone{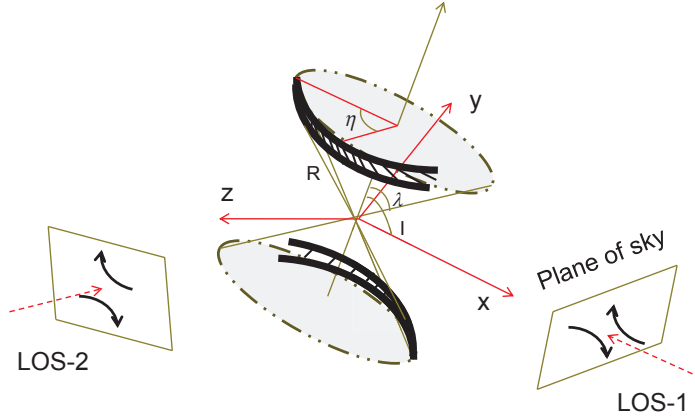}
\caption{\small  Schematic illustration showing an X-shaped  source
under the non-ballistic precession scenario. Different precession
states of the X-sources  are equivalent to the observation of
one source at different view angles as LOS$_1$ and  LOS$_2$, and hence different morphologies.
\label{carton} }
\end{figure}

This model can be described by two simple geometric equations.
Projecting a knot, $i$, with knot-core separation, $R^i$, to the
coordinate system $x-y-z$, we have,
\begin{eqnarray}\label{lkx}
 R^i_{x} &=& R^i[\sin\lambda\sin I\cos\eta^i +\cos\lambda\cos I ] \,, \nonumber\\
 R^i_{y} &=& R^i[\sin\lambda\sin\eta^i ]\,, \nonumber\\
 R^i_{z} &=& R^i[\cos\lambda\sin I -\sin\lambda\cos I\cos\eta^i ]
\,,
\end{eqnarray}
where $\lambda$, $I$ , $\eta$, and  $R$, represent the opening angle of the
precession cone, the inclination angle between the jet rotation axis
and the line of sight (LOS), the precession phase, and  the knot-core distance respectively, as shown in
Fig.~$\ref{carton}$.

\clearpage

In Eq.~($\ref{lkx}$) the $x$-axis is towards the observer. Rotating around the
$x$-axis for angle $\xi$, so that the new $y$-axis ($\Delta\delta$)
will point north, and the new $z$-axis ($\Delta\alpha$) will
point east. Therefore, the coordinate of a feature in the plane of sky is given,
\begin{eqnarray}\label{lk}
R^i_{\alpha} &=& R^i[\sin\lambda\sin\eta^i ]\sin\xi+R^i[\cos\lambda\sin I -\sin\lambda\cos I\cos\eta^i   ]\cos\xi   \,, \nonumber\\
R^i_{\delta} &=& R^i[\sin\lambda\sin\eta^i ]\cos\xi-R^i[\cos\lambda\sin I -\sin\lambda\cos I\cos\eta^i ]\sin\xi
\,,
\end{eqnarray}
Eq.~($\ref{lkx}$) predicts an  ellipse, the shape of which is determined by four parameters,  $\xi$, $\lambda$, $I$, and  $R$.
Searching in the parameter space of them  as shown in the brackets of Table~1,  the best combination of parameters corresponding to the solid ellipses in Fig.~$\ref{morphology}$ can be found.

Once the best fit ellipse for a morphology is found, the
precession phase of a knot, $\eta^i=\Omega t+\eta^i_0$, can be given, where $\Omega$ is  the precession velocity of a jet, and $\eta^i_0$ is the initial phase of a knot.
However, the
precession time, $t$, and the precession velocity, $\Omega$, can not
be separated from $\eta^i$. In other words,
$\Omega$ can not be obtained by such a 2-dimensional morphology fitting
alone.

Fortunately, a simple constraint can be found  to split $\Omega$
and $t$, and therefore allows us to determine the
precession period by 3-dimension kinematic. The time of emission of a photon  from a knot,
 $t_{emit}^i$,  can be  measured in the reference frame  at rest to
the core of an X-shaped source. This photon  can  reach an observer on Earth
at, $t_{arr}=t_{emit}^i+d/c-x^i/c$, where $d$ is the
core-observer distance, and $x^i$, which is equivalent to $ R^i_{x}$ of
Eq.~($\ref{lkx}$) is the projection of the knot-core separation onto
LOS. For simplicity, we can define a time, $t_a=t_{arr}-d/c$, so
that the time taken for a photon from a knot to the observer can be
represented simply by, $t_{a}=t_{emit}^i-x^i/c$.

Also measured in the reference frame at rest to the core of an
X-shaped source, the precession time of a knot can be synchronized
 to the emitting time,  $t^i=t_{emit}^i$, where $t^i$
is given by $\eta^i=\Omega t^i+\eta^i_0$. Thus, the condition of
observing the two signals from two opposite lobes ($i=1,2$ denote the north and south lobe respectively) at the same time becomes:
\begin{equation}
\label{timedelay} t^i-x^i/c= t_{a} \,. \ \
\end{equation}
Certain precession phases result in, $ t^i-x^i/c > t_{a}$, which
means that the signal has not arrived to the observer yet, and is
hence unobservable, whereas, $ t^i-x^i/c < t_{a}$ means that the
received signal is from a knot that has been at its emitting site
for a period of time. If it has afterglow emission that is above
the threshold of detection, then it is still observable.

Therefore, a knot which is unobservable in the case of zero cooling
time, $ t^i-x^i/c < t_{a}$, becomes  detectable provided the
emissivity of the knot is above the threshold of detection in the
cooling time, $t^i_c$. Such an emission reaches  the Earth at:
\begin{equation}\label{cooling}
t^i_{c}+ t^i- x^i/c=t_a \,.
\end{equation}
Consequently, with larger and larger
separation of the active lobes, a knot in the wings corresponds to
larger and larger $t^i_{c}$, so that it appears more
diffused and faint until it becomes unobservable.

Limited by the expansion speeds of low-luminosity FR-II
sources, the advance speeds of  X-sources are likely not greater
than $0.04$c~\citep{DT02}. Such non-relativistic speeds correspond
to a negligible Doppler boosting effect, so that the flux of a knot
depends primarily on the emissivity. Hence, the precession period at
the frame of the source, $P_p$, and at the Earth, $P_p^{obs}$,  are
related by $P_p(1+z)=P_p^{obs}$ , where $z$ is the red-shift of the
source.

Multiplying the precession velocity,  $\Omega$, at the two sides of
 Eq.~($\ref{timedelay}$), we have,
\begin{equation}
\label{timedelayphase} \eta^i-\eta^i_0-\Omega x^i/c=\eta_{a} \,, \ \
\end{equation}
where $\eta_a=\Omega t_a$. Apparently,  an X-source is observed when  the photons from the lobes
arrive at the Earth simultaneously. At this moment the initial phases of the two lobes can be treated as,
$\eta^1_0=\eta^2_0$, without losing generality. Thus $\eta^i_0$ of  Eq.~($\ref{timedelayphase}$) can be canceled.

The $\eta^i$ and $x^i$ of the north and south components
can be obtained by fitting the 2-dimensional
X-shaped morphology. The process is simply  putting the geometrical
parameters of Table~1 (except $\Omega$ and $\Omega^{obs}$) into
Eq.~($\ref{lk}$), then find the best-fit parameters through  minimizing
the sum of the square of the residuals of the predicted ellipse from the observed morphologies.
The pair ellipses (with same group of geometric parameters) for each source are required to  fit 7-8 components represented by the red dots in Fig.~$\ref{morphology}$. The solid ellipses of Fig.~$\ref{morphology}$ represent the best fit ones corresponding to parameters shown in Table~1, and the dashed ellipses correspond to $20\%$ increase in  parameters, $I$, $\lambda$, and $\xi$ (while  holding  others constant) for
 3C52, 3C223.1 and   4C12.03  respectively. Hence, the role  of these parameters in the formation of an ellipse is exhibited,
i.e.,  for 4C12.03, the 20$\%$ increase in $\xi$ from its best value changes not only the shape of the dashed ellipse, but also its position, in which case, the morphology cannot be fitted, no matter what  other parameters are.

Although making an ellipse through 7-8 points  in the morphology strongly constrains the fitting parameters, we did find that the morphology of  3C223.1 can be fitted by other combination of parameters, i.e., $R$  approximately twice and $I$ half of the corresponding parameters of Table~1. However, such a solution predicts much larger discrepancy in  propagation time between the Northwest(NW) and Southeast(SE) structure than that of Table~1, which is contradict to the nearly symmetric structure of 3C223.1, as analyzed in Section~3. It is thus excluded.

Differentiating Eq.~($\ref{lk}$)  one  has, $\Delta R^i_{\kappa}=\sum f_j\Delta\sigma_j$, where $\sigma_j$($j$ from 1 to 4), denotes $\lambda$, $I$, $\xi$ and $R$ respectively, and $f_j$ correspond to their partial differentiations respectively. If  $\Delta R^i_{\kappa}$, where $\kappa$ represents $\alpha$ and $\delta$ of  Eq.~($\ref{lk}$), could be as large as the size of a lobe as shown by red circles in Fig.~$\ref{morphology}$ (which is  attributed to the error of the precession phase of a lobe), then the errors of  $\sigma_j$ can be obtained by solving four equations,  $(\Delta R_{\kappa}^i)^2=\sum f_j^2(\Delta\sigma_j)^2$,  corresponding to two lobes, $i=1,2$ (the north and south),  of a source. This gives conservative errors (up to $43\%$) to the best fit parameters of  Table~1, which considerably exceed the $20\%$ parameter errors corresponding to the deviation between the solid and dashed ellipses shown in   Fig.~$\ref{morphology}$.

Although the 2-dimension morphology fitting perpendicular to
the LOS can give $\eta^i$ and $x^i$, the most interesting
parameter, $\Omega$,  can not be extracted from
Eq.~($\ref{timedelayphase}$). Since the NW and SE structure
indicates that photons from them must arrive at
the Earth simultaneously, this can be treated as another constraint (1-dimension),
along the LOS.  The two active lobes(the north and south lobes) with the
shortest cooling time always satisfy $t^1-x^1/c= t^2-x^2/c$, no
matter if $t^1\approx t^2$ ($x^1\approx x^2$), or if these values
differ largely. Hence  the kinematics of the two active lobes
of an X-source reads, $\eta^1-\Omega x^1/c=\eta^2-\Omega x^2/c$, which includes 3-dimension constraints, and from which  $\Omega$ can be extracted.

\begin{figure}
\epsscale{0.7} \plotone{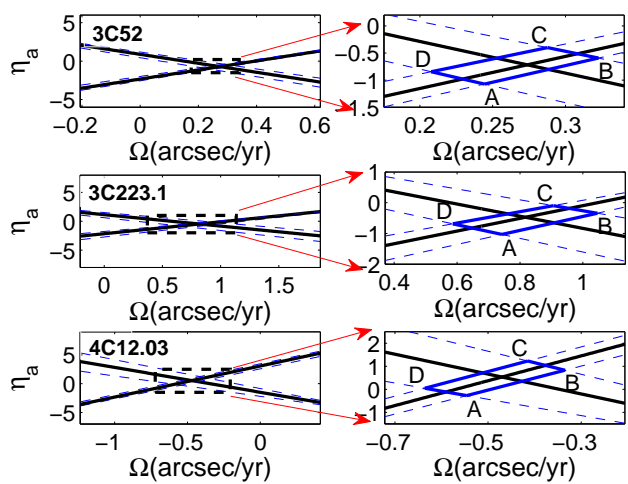}
\caption{\small The determination of the precession period of three X-sources.
The two lines that cross at ($\Omega, \eta_{a}$),
represent the same time of arrival of photons from the north and south
lobes. The dashed lines correspond to uncertainty in $\eta^i $ and $
x^i$, estimated by the size of the lobes. The jet precession
velocity of a source, $\Omega$, is thus constrained in the small
regions, ABCD. \label{cross} }
\end{figure}

To obtain $\Omega$ and its error from Eq.~($\ref{timedelayphase}$), both $\Omega$ and $\eta_{a}$ can be ordered as variables.
Thus, Eq.~($\ref{timedelayphase}$) corresponds
to two lines ($i=1,2$), the cross point of  which, ($\Omega,\eta_{a}$), represents that the two signals  arrive on the Earth simultaneously, as shown in Fig~$\ref{cross}$.
With $\eta^i$ and $x^i$ of the  north and
south lobes obtained through morphology fitting (2-dimension), and utilize
the constraint on photon arrival time (1-dimension), the cross point ($\Omega, \eta_{a}$) can be
determined by  Eq.~($\ref{timedelayphase}$), through which the precession
velocity, $\Omega$, can be constrained into a small range
elegantly, as shown in Table~1. The error of $\Omega$ is determined by the errors of  $\eta^i$ and $x^i$,   where  $\Delta x^i$ can be given by the error  propagation of parameters in Table~1 via  Eq.~($\ref{lkx}$), and  $\Delta\eta^i$ is assumed to correspond to the ``radius" of a lobe.

\section{Asymmetry}

The fitting  parameters of Table~1 indicates that the phase discrepancy
of the two lobes are $\delta\eta^{12}=185^{\circ}$ and
$\delta\eta^{12}=189^{\circ}$ for 3C52 and 3C223.1 respectively.
The $\approx 180^{\circ}$ phase discrepancy corresponds to the observation of an X source through the LOS$_2$ of
Fig~$\ref{carton}$.
Correspondingly,
$t^1-x^1/c=t^2-x^2/c$ is satisfied in the case $x^1\approx x^2$ and
$t^1\approx t^2$. As the time $t^1$ and  $t^2$ increase,
the values of $x^1$ and  $x^2$ change similarly. Therefore,
the emission from the  lobes and wings can arrive the Earth at approximately the same time.
Consequently, the SE and NW structures of these two sources  appear comparable in length and  size.

In contrast to these two X-sources, the north and south lobes of 4C12.03
correspond to a phase discrepancy of
$\delta\eta^{12}=\eta^1-\eta^2\approx -270^{\circ}$ instead of $\approx 180^{\circ}$, by the fitting parameters of Table~1.
This means that the photons from the active south lobe, which
should differ by approximately $\approx180^{\circ}$ to the precession phase
of the north lobe, have not reached the observer on the Earth yet ($
t^2-x^2/c>t_a$), although such a lobe is observable
at the core of this X-source. This corresponds to the observation of such a source through the LOS$_1$ of
Fig~$\ref{carton}$. Therefore,
$t^1-x^1/c=t^2-x^2/c$ is satisfied in the case $x^1$ and $x^2$ ($t^1$ and $t^2$ also) differ significantly. Such a discrepancy in $x^1$ and  $x^2$  results in  a Southwest(SW) pattern compared to that of Northeast(NE). Hence, the significant asymmetry of 4C12.03 is well understood.

The  north  and   south lobes  of 4C12.03 obviously differ in size.
By the fitting parameters of Table~1, the
active north lobe posses the maximum $t^1$ and minimum $\eta^1$ (due to
$\Omega<0$ in 4C12.03), which means that knots of $t>t^1$ and
$\eta<\eta^1$  don't exist in the NE pattern at all. This is in agreement with
both observers on the Earth and at the core of this
X-source. Consequently, the emission of this lobe region can only be
extended by the past (cooled) components, with $t\leq t^1$ and $\eta
\geq \eta^1$.

Contrarily, for the SW pattern, its emission can be extended by both
the past components with $t^2-x^2/c\leq t_a$, and by some ``future"
components with $t^2-x^2/c> t_a$. Because components with $t>t^2$
and $\eta<\eta^2$ do exist near the ``south lobe" (which would be
observable at the core of this X-source), and the cooling emission
of such ``future" knots can contribute to the brightness of the ``south
lobe" as well by $t^2_c+t^2-x_{f}^2/c= t_a$ (where $x_{f}$ denotes
the ``future" emission site). Therefore, the one way extension of
the NE lobe and the two way extension of the ``SW lobe" lead to a larger
 ``south lobe" than that of the NE one.

\begin{figure}
\epsscale{0.42} \plotone{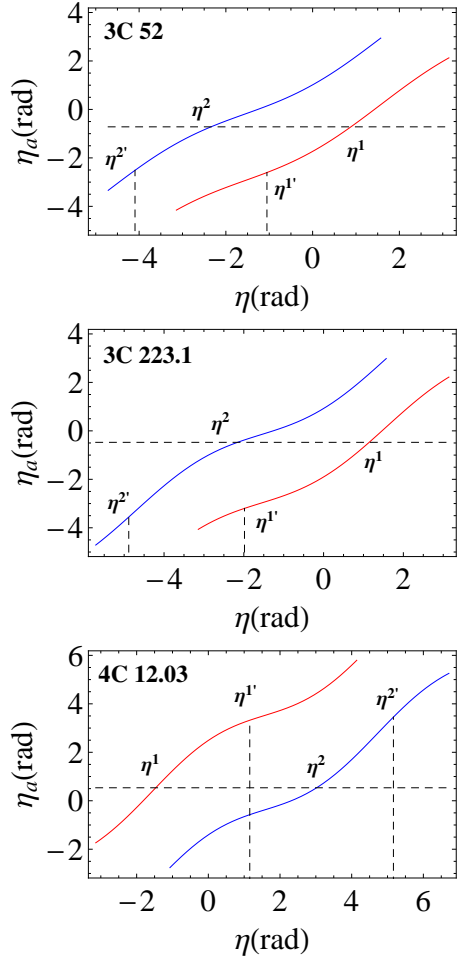}
\caption{\small The precession phase versus the phase of arrival
time, with $\Omega=const$.  The cross points of the horizontal line
with two curves, $\eta^1$ and $\eta^2$  corresponds to  the precession phases of
the north and south lobe respectively. The cross points of the curves with  the
vertical lines ($\eta^{i\prime}$) correspond to the precession phase
of the tail of wings.
 \label{phasebined} }
\end{figure}

To analyze the fine asymmetry in these X-sources,
Fig~$\ref{phasebined}$ is introduced, which is obtained as follows.
Putting the obtained $\eta^i$ and $x^i$ through
the fitting of morphologies of three sources
of Fig~$\ref{morphology}$ into Eq.~($\ref{timedelayphase}$), then ordering
$\Omega=const$,  the phase corresponding to the time of
arrival, $\eta_a$, versus the phase of precession, $\eta$, can be
obtained as  shown in Fig~$\ref{phasebined}$, which  is actually
the evolution of $t_a$ and $t$ of the two active lobes.

As shown in Fig~$\ref{phasebined}$,  each source has its own
$\eta-\eta_a$ curve and  horizontal dashed line, $\eta_a$, which cross
with a misalignment angle. In both 3C52 and 3C223.1, the
misalignment angles of the south lobe are smaller than those of the north
lobe. Moreover, the discrepancy in the misalignment angles of the
north and south lobe is more obvious in 3C223.1 than in 3C 52. This
explains the deviation in surface brightness of the NW and SE lobe
regions in 3C223.1, as shown in Fig~$\ref{morphology}$. Because
the region near the south lobe is closer to the horizontal line,
$\eta_a$, than that of the north one, which corresponds to a shorter cooling
time. Hence the linking of the south lobe and its neighboring wing
region has higher surface brightness than those of the north one.

\section{Discussion}

By the fitting of Fig~($\ref{morphology}$)  and parameters of
Table~1, precessing across the lobe-wing region takes 1.5 Myr, 0.8
Myr and 1.4 Myr  for  3C52, 3C223.1 and 4C12.03 (NE pattern)
respectively, the time scale of which corresponds to the cooling
time discrepancy between the active lobe and the tail of wing of
these sources. Interestingly, the time scale is consistent with the firm
upper limits on the particle ages of 34 Myr for 3C223.1, and the
estimation of the time scale of reorientation of jet axis of no more
than a few Myr based on  spectral gradient~\citep{DT02}.

Beside the temporary processes such as  merger of SMBBH or
disk instability~\citep{ME02,DT02}, the change of jet orientation
displayed in X-shaped morphology can also be originated from binary effect~\citep{Rees80}, which can be
either relativistic geodetic  precession or Newtonian-driven
jet precession~\citep{Katz97}.
A  SMBBH system with orbital period of 20 years, and with a typical X-source black hole mass of $5\times10^8M_{\bigodot}$~\citep{Bonn} and companion mass of $1\times10^7M_{\bigodot}$, predicts a precession period of 0.6Myr by the geodetic effect; and a precession period of 1.0Myr (with disk radius of 10 schwarzschild radius) by  the  Newtonian driven effect. Consequently, the $\sim$Myr time scale displayed in the three X-sources can be well interpreted by either of the two binary  effects.  This provides another evidence to the link between X-sources and SMBBHs. Moreover, the correlation of black hole mass with X-ray
luminosity and radio luminosity\citep{Merloni03,Falcke04}; as well as  characteristic
time-scale of the X-ray variability finding in AGNs and  X-ray binaries\citep{McHardy06,Kording07,Mirabel09}
suggests that black hole physics likely scales with mass. The  $\sim$Myr precession periods of the three X-sources revealed here further support such a correlation with black hole mass.

In the non-ballistic precession model, the diversity of  morphology of X-sources
can be simply understood by the received photons from  emission sites at different precession cones.
The technique is applicable to other  X-sources, especially the X-shaped candidates~\citep{Cheung07} in the future.

\section{Acknowledgments}
We thank  Y.C. Zou  and Z.Q. Shen for helpful discussions. This research is supported by the
National Natural Science Foundation of China, under grant
NSFC10778712.

\begin{table}
\begin{center}
\caption{\bf The parameters extracted by morphology fitting and
3-dimension kinematics.}
\resizebox{18cm}{!} {
\begin{tabular}{ccccccccc}
\hline \hline

Source & $\xi(0,2\pi)$ & $\eta^{1}(0,2\pi) $ & $\eta^{2}(0,2\pi) $ & $I(0,\pi/2)$  & $\lambda^{1}(0,\pi/2)$
  &  $R(0,2)$ & $\Omega$(arcsec/yr) &  $\Omega^{obs}$(arcsec/yr)\\

3C52 & $5.2\pm 0.14$ & $0.88\pm0.44$  & $-2.4\pm0.34$ & $0.52\pm0.065$ &
$0.43\pm0.040$ & $0.40\pm0.093$ & $0.27\pm 0.062$ & $0.21\pm 0.048$ \\

3C223.1 & $5.1\pm0.061$ & $1.1\pm0.50$ & $-2.2\pm0.41$  & $0.62\pm0.23$ &
$0.45\pm0.077$ & $0.15\pm0.021$ & $0.83\pm 0.23$ & $0.75\pm 0.21$\\

4C12.03 & $3.9\pm0.046$ & $-1.5\pm0.66$ & $ 3.1\pm0.50$  & $0.65\pm0.15$
& $0.35\pm0.079$ & $0.35\pm0.11$ & $-0.47\pm0.17$ & $-0.41\pm 0.15$\\
\hline \hline
\end{tabular}
}
\end{center}
{\small The precession velocities, $\Omega$, in the frame of the
source, are inferred from Fig~$\ref{cross}$, which is related to
the frame of the Earth by  $\Omega=\Omega^{obs} (1+z)$.  The
rest are precession parameters obtained by the fitting of
Fig~$\ref{morphology}$. Angular parameters  are in rad, in which
$\eta^{1} $ and $\eta^{2}$ represent the phase of the north and south lobe
respectively. The opening angle of precession cone, $\lambda^{1}$,
corresponds to the NW(NE for 4C12.03) pattern, and the  opposite pattern is fitted by
$\lambda^{2}=\pi-\lambda^{1}$. The distance $R$ is in Mpc, which
 is obtained by multiplying the arcsec value (through morphology fitting) by the
angular size distance (through the cosmology parameters and the measured redshift).
}
\end{table}

\end{document}